\documentclass[12pt]{article}

\usepackage[active]{srcltx}
\usepackage{latexsym}
\usepackage{amsmath}
\usepackage{amsfonts}
\usepackage{mathrsfs}
\usepackage{amssymb}
\usepackage{dsfont}
\usepackage{amscd}
\usepackage{cite}
\usepackage{verbatim}
\usepackage{xcolor}
 \csname
@addtoreset\endcsname{equation}{section}

\usepackage{lipsum}
\usepackage{xargs}
\usepackage[normalem]{ulem}
\allowdisplaybreaks

\usepackage[colorinlistoftodos,prependcaption,textsize=tiny]{todonotes}
\newcommandx{\unsure}[2][1=]{\todo[linecolor=red,backgroundcolor=red!25,bordercolor=red,#1]{#2}}
\newcommandx{\change}[2][1=]{\todo[linecolor=blue,backgroundcolor=blue!25,bordercolor=blue,#1]{#2}}
\newcommandx{\Sinfo}[1]{\todo[backgroundcolor=red!25,bordercolor=red,noline]{S:#1}}
\newcommandx{\Winfo}[1]{\todo[backgroundcolor=blue!25,bordercolor=blue,noline]{W:#1}}

\def\p{\partial}

\def\a{\alpha}
\def\b{\beta}
\def\g{\gamma}

\def\d{\delta}

\def\l{\lambda}

\def\m{\mu}

\def\r{\rho}

\def\s{\sigma}

\def\t{\tau}
\def\u{\upsilon}

\def\be{\begin{equation}}
\def\ee{\end{equation}}
\def\bea{\begin{eqnarray}}
\def\eea{\end{eqnarray}}
\def\ba{\begin{array}}
\def\ea{\end{array}}

\def\12{\frac{1}{2}}

\def\eNm1{\overset{\scriptscriptstyle{(N-1)}}{e}}

\def\N{\mathbb{N}}

\begin{document}

\begin{flushright}
\end{flushright}

\vspace{25pt}

\begin{center}


{\LARGE \bf Osborn Equation and Irrelevant Operators\\
\vskip.5cm
}

\vspace{25pt}
{\large Adam Schwimmer$^{a}$ and Stefan~Theisen$^{b}$}

\vspace{10pt}
{\sl\small
$^a${\it  Weizmann Institute of Science, Rehovot 76100, Israel}
}

\vspace{10pt}
{\sl\small
$^b${\it Max-Planck-Insitut f\"ur Gravitationsphysik, Albert-Einstein-Institut, 14476 Golm, Germany}

}

\vspace{70pt} {\sc\large Abstract}\end{center}


The structure of the Osborn (``Local Renormalization Group")
Equation in the presence of integer dimensional irrelevant operators is studied.
We argue that the consistency of the anomalous part of the generating functional requires a beta-function for the metric.
The modified form  of  the Weyl anomalies is calculated.

\newpage




\section{Introduction}

Integer dimensional primary operators in Conformal Field Theories (CFT) give rise to conformal
anomalies \cite{Duff}.
While at separate space-time points the correlators of these operators are well defined, at
coincident points the singularity needs a regulator. As a consequence there is an irremovable violation
of the conformal Ward identities. The singularities involved are typically logarithmic,
leading to the so-called ``type B" conformal anomalies \cite{DDI}. 
In particular  situations involving identities valid in special space-time
dimensions, the logarithms are converted into integer power singularities without a scale
(``type A" conformal anomalies) \cite{DS}.

Higher order correlators of integer dimensional operators produce increasing powers of logarithms. The recursive
nature of the correlators related to their analytic structure in momentum space reduce the higher powers of logarithms to
the independent, universal information contained in the single logarithm  term in each correlator.

This structure is summarized in the Osborn Equation. While the equation can be used to describe the
perturbative renormalization group flow in a CFT perturbed by relevant operators,
in the present paper we will use it strictly
in the CFT: the space-time dependent couplings will be considered as sources of the primary operators. The coefficients
of the Taylor expansion of the partition function around zero sources are the CFT correlators.

We give  now a brief description of the Osborn Equation \cite{Osborn}.\footnote{For further discussions and
applications see also \cite{JO1,JO2,Baume}.}
Consider in a CFT relevant and marginal primaries
${\cal O}_i$ with integral dimensions $\Delta_i$. The operators ${\cal O}_i$ are coupled to
space-time dependent sources $J_i(x)$. In the list of primary operators we include the energy-momentum
tensor $T_{\mu\nu}(x)$ coupled to the background metric $g_{\mu\nu}(x)$.
All other operators which we  will consider in this work are scalars.
The connected generating functional $W$ depends on $J_i(x)$ and $g_{\mu\nu}(x)$.
The local Weyl variation operator acts on $W$ as
\begin{equation}\label{deltaW}
\delta_\s W=\int\,d^d x\left\{2\,\s(x)g_{\mu\nu}(x){\d\over\d g_{\mu\nu}(x)}
+\sum_i\beta_i\big(\{J_j\},g_{\mu\nu};\s \big){\d\over\d J_i(x)}\right\}W
\end{equation}
where the first term represents the canonical Weyl transformation of the metric and the second term involves
the local ``beta-functions" of the sources (couplings) linear in the infinitesimal Weyl parameter $\s(x)$ and
its derivatives.

We remark that the non-classical part of the beta-function, i.e. the part which contains information 
about logarithms in correlation functions,  arises from the action of the Weyl transformation on the anomalous part 
of the generating functional. Equivalently, the correlation functions satisfy the  
$SO(d,2)$ (the conformal group in $d$ dimensions) Ward-identities, as long as all operators are 
inserted at separate points.

The Osborn Equation states that the Weyl variation of $W$ gives the Weyl anomalies
\begin{equation}\label{Osborneq}
\delta_\s W=\int d^d x\sqrt{g}\,\s(x){\cal A}(\{J_i\},g)
\end{equation}
where ${\cal A}$ is the local anomaly polynomial.
The generating functional can be chosen to respect diffeomorphism symmetry. Then
there are two cohomological structures underlying  the Osborn Equation which
are related to the realization of the abelian  Weyl symmetry:

\begin{enumerate}

\item Since the Weyl transformation should obey
\begin{equation}\label{Weylgroup}
[\delta_{\s_1},\delta_{\s_2}]=0
\end{equation}
the transformation of the sources implied by \eqref{deltaW}
\begin{subequations}\label{deltagJ}
\begin{align}
\delta_\s g_{\mu\nu}&\equiv 2\,\s\, g_{\mu\nu}\label{deltag}\\
\delta_\s J_i&\equiv\beta_i\big(\{J_j\},g_{\mu\nu};\s\big)\label{deltaJ}
\end{align}
\end{subequations}
should obey the integrability condition following from \eqref{Weylgroup}. While \eqref{deltag} obviously obeys it, for
\eqref{deltaJ} we get the non-trivial constraint
\begin{equation}\label{WZbeta}
\delta_{\s_2}\b_i\big(\{J_j\},g_{\mu\nu};\s_1\big)=\delta_{\s_1}\b_i\big(\{J_j\},g_{\mu\nu};\s_2\big)
\end{equation}
where the second variations are calculated using \eqref{deltagJ}.

The integrability conditions have trivial solutions corresponding to local changes of variables of the sources
\begin{equation}\label{Jprime}
J_i'(x)\equiv\gamma_i\big(\{J_j\},g_{\mu\nu}\big)
\end{equation}
such that $J_i'(x)$ transforms just following the dimension, i.e. without a non-trivial
beta-function:
\begin{equation}\label{deltaJprime}
\delta_\s J_i'(x)=\s(x)\big(\Delta_i-d\big)J_i'(x)
\end{equation}
The solutions of \eqref{WZbeta} modulo \eqref{Jprime} define the cohomologically nontrivial
beta-functions.

Once these non-trivial beta-functions are found, the Weyl anomaly, i.e. the r.h.s. of
eq.\eqref{Osborneq}, is the solution of a second cohomology problem:

\item The anomaly should obey
\begin{equation}\label{WZanomaly}
\delta_{\s_2}\int d^d x\sqrt{g}\,\s_1(x){\cal A}\big(\{J_i\},g_{\mu\nu}\big)
=\delta_{\s_1}\int d^d x\sqrt{g}\,\s_2(x){\cal A}\big(\{J_i\},g_{\mu\nu}\big)
\end{equation}
where the variations are calculated using \eqref{deltag} and \eqref{deltaJ}. This is the Wess-Zumino
(WZ) consistency condition.
Cohomologically nontrivial
Weyl anomalies are solutions of \eqref{WZanomaly} modulo variations of local diffeo-invariant functionals
of $\{J_i\}$ and $g_{\mu\nu}$.

\end{enumerate}

When the beta-function reduces to the classical term \eqref{deltaJprime},
the Osborn Equation produces the ordinary Weyl anomalies related to single
logarithms in the generating functional.
The universal information contained in the beta-function is the coefficient of the single logarithm in
the correlator corresponding to each power of the source in the expansion.
The presence of the nontrivial part of the beta-function is related to multiple logarithms
in the generating functional: due to the recursive relations incorporated into the
equation \eqref{Osborneq} the logarithms in the left hand side of the equation are cancelled
producing at the end the anomaly.

The role of the beta-function can be made explicit in the lowest non-trivial order.
Consider the correlator of three integer dimensional operators ${\cal O}_i , {\cal O}_j ,{\cal O}_k$ which has a double logarithm
corresponding to a nonvanishing structure constant (the simplest realization of this situation is for three
marginal but not truly marginal operators). This means that at least for one choice of the indices the OPE
(operator product expansion) of two  operators contains the third one with a logarithmic singularity:
\begin{equation}\label{OPE1}
{\cal O}_i(x) {\cal O}_j(y)=c_{ijk} |x-y|^{\Delta_k-\Delta_i-\Delta_j} {\cal O}_k(y)
\end{equation}
Together with the logarithm in the $\langle {\cal O}_k(y) {\cal O}_k(z)\rangle$  correlator this produces the
double logarithm in the three point function  or equivalently in the appropriate term in the
generating functional:\footnote{the exact position of the double logarithms is not essential for the following argument.}
\begin{equation}\label{logs}
W=... +c_{ijk}J_i J_j (\log(\Box/\mu^2))^2 J_k+ ..+J_k \log(\Box/\mu^2)J_k+...
\end{equation}
where we made explicit also the contribution of the ${\cal O}_k$ two point function and we left out
the powers of $\Box$ implied by the dimensions of the operators.
The singularity in the OPE is translated into the first nontrivial term in the beta-function of the source $J_k$:
\begin{equation}\label{beta}
\delta_\s J_k\equiv \beta_k=\sigma c_{ijk}J_i J_k+...
\end{equation}
Now we can follow the cancellation of the single logarithm in \eqref{Osborneq} corresponding to \eqref{logs}:
the action of the Weyl variation of the metric on the covariant Laplacian in the first term cancels the
variation of the second term when acted upon by  $\beta_k \frac{\delta}{\delta J_k}$ .

The recursion relations which are underlying the  perturbative validity of the Osborn Equation are a
consequence of the analytic structure of the CFT: in momentum space the correlators obey
(subtracted) dispersion relations and a finite number of subtractions depending on the dimensions
of the operators control all correlators. This is of course valid provided the operators
are relevant ($\Delta_i<d$) or marginal ($\Delta_i=d$).

For integer dimension irrelevant operators ($\Delta_i>d$) the number of subtractions needed increases
with the order of the correlator, but we expect that for any finite number of irrelevant operators
the general structure of the Osborn Equation will be still valid since one needs a finite number of subtractions
to control the logarithms.

We will see, however, that in the presence of irrelevant operators a rather basic feature of eq.\eqref{deltaW} is
changed: we have to introduce a non-trivial beta-function also for the transformation of
$g_{\mu\nu}$. This changes the definition of the Weyl transformations but it is necessary since
correlators of irrelevant operators with energy-momentum tensors
acquire an increasing  number of logarithms. This  feature becomes evident already for the terms in the
generating functional involving only two irrelevant operators when as for relevant or marginal operators the beta-function
does not contribute. The sources transform just following their dimension and the Osborn Equation should produce a
unique local Weyl anomaly.
There is, however, a general proof that such a local expression with the required
cohomological properties cannot exist for two irrelevant operators. We correlate this feature with the existence of
multiple logarithms in correlators involving energy-momentum tensors and we argue that a ``metric beta-function"
should be introduced, which modifies the transformation
rule of the metric under a Weyl transformation. 
The non-classical contribution contains information about logarithms and acts on the anomalous part of the 
effective action, in analogy to the beta-function of the scalar operators which we have discussed before.
We analyze in detail the consistency of this assumption in lowest non-trivial
order through the OPE involving the energy-momentum tensor and through a general cohomological analysis.
We discuss the physical consequences of this generalization, in particular the implications for the RG transformation
properties of the geometry dependence  when the CFT is formulated on a compact manifold.

In Section 2 we review the Weyl anomaly involving two relevant or marginal operators and the mathematical results
determining its form. Similar expressions for two irrelevant operators are proven not to exist. We study the OPE
involving the energy-momentum tensor and two irrelevant operators generating a metric beta-function to lowest
non-trivial order.

In Section 3 we examine in detail the cohomological structure of the metric beta-function and corresponding anomaly in $d=4$.
The explicit expression is worked out for the lowest irrelevant dimension, i.e. $\Delta=5$.

In Section 4 we discuss the peculiar features produced by irrelevant operators in $d=2$, in particular the
interplay with type A Weyl anomalies.

In the Conclusions we discuss the general structure of Weyl anomalies involving  an arbitrary maximal number of
irrelevant operators and energy-momentum tensors. We also point out the implications for the dependence of the partition
function on the metric when the theory is formulated on a compact manifold: in the presence of an irrelevant deformation
the background geometry is ``running" under a RG transformation.
The relevance of the irrelevant operator anomalies for the structure of the chiral ring in supersymmetric gauge theories
is indicated.

\section{Anomalies and beta-functions}

Given a CFT in $d$ dimensions, consider the two-point function of scalar primary operators of dimension $\Delta$.
With a suitable normalization, conformal invariance fixes it to be
\begin{equation}\label{OO}
\langle {\cal O}(x)\,{\cal O}(0)\rangle={N\over |x|^{2\Delta}}
\end{equation}
For $\Delta-d/2=n\in\N_0$ this is singular, i.e. it has no Fourier transform. It needs regularization and
this necessarily introduces a scale.
This is most easily done in momentum space where one finds
\begin{equation}\label{OO}
\langle{\cal O}(p)\,{\cal O}(-p)\rangle=(-1)^{n+1}{N\,\pi^{d/2}\over 2^{2n}\Gamma(n+1)\Gamma(n+{d\over2})}
 p^{2n} \Big(\log(p^2/\mu^2)+c_{n,d}\Big)
\end{equation}
where the constants $c_{n,d}$ are scheme dependent; they change if one rescales  $\mu$.
The explicit scale dependence signals a type B conformal anomaly: the two-point function transforms
inhomogeneously under dilatations, i.e. rescaling of the coordinates
$x^\mu\to e^\lambda x^\mu$ or, under rescaling the momenta $p_\mu\to e^{-\l}p_\mu$.
Equivalently
\begin{equation}
\mu{d\over d\mu}\langle{\cal O}(p)\,{\cal O}(-p)\rangle=(-1)^n {N\pi^{d/2}\over 2^{2n-1}\Gamma(n+1)\Gamma(n+{d\over2})}p^{2n}
\end{equation}
In position space this translates into
\begin{equation}\label{Anomaly2ptx}
\mu{d\over d\mu}\langle{\cal O}(x)\,{\cal O}(0)\rangle\,\propto\,\square^n\delta(x)
\end{equation}
Another way to express the anomaly, which was already used in the Introduction,
is based on the fact that Weyl invariance in curved space-time implies
conformal invariance in Minkowski space. One introduces space-time dependent sources
for ${\cal O}$ and the energy-momentum tensor, $J(x)$ and $g_{\mu\nu}(x)$.
The correlation function \eqref{OO}  then translates into a term in the generating functional $W$ proportional to
\begin{equation}\label{Anomaly2ptp}
\int d^d x \sqrt{g(x)}\,J(x)\square^n\log(\square/\mu^2)J(x)
\end{equation}
and the anomaly as the non-invariance of $W$ under Weyl rescaling the sources, i.e.
\begin{equation}\label{dWn}
\d_\s W=\int d^d x\,\s\,J\,\square^n J
\end{equation}
We have dropped terms which vanish in flat space-time.
In this basic form the anomaly expresses  a violation of the Ward identity which relates the correlator
$\langle T_{\mu\nu} \cal O\cal O \rangle$ to the two point function $\langle \cal O \cal O\rangle$.

Due to the nonabelian nature of the diffeomorphism group the Ward identity is extended to correlators with
any number of energy-momentum tensors and the anomaly expression contains
metric dependent terms constrained by diffeomorphism invariance. In the general framework of the Osborn Equation this
corresponds to the term in the expansion of the effective action which is quadratic in the sources.
To this quadratic order the non-trivial part of the beta-function of the sources cannot contribute.
The construction of these terms, which depend on
the space-time curvature, is well known in the case of relevant and marginal operators, i.e. for $\Delta\leq d$.
Of course we assume $\Delta-{d\over2}=0,1,2\dots$ as otherwise there is no anomaly. We will now review
this case. We start with  a marginal operator and its source $J$.
$J$ is invariant under Weyl rescaling and the anomaly is
\begin{equation}
\delta_\s W=\int d^d x\,\sqrt{g}\, \s\,J\Delta_c J
\end{equation}
where in $d=2$, $\Delta_c=\square$, and in $d=4$,
\begin{equation}
\Delta_c=\square^2+2\nabla_\mu\big(R^{\mu\nu}-\textstyle{1\over3}g^{\mu\nu}R\big)\nabla_\nu
\end{equation}
Here $\square$ is the covariant Laplacian and $R$ the curvature.
They transform as $\Delta_c\to e^{-d\s}\Delta_c$. In $d=4$, $\Delta_c$ is known as the Fradkin-Tseytlin-Riegert
operator in the physics literature and as  the (critical) Paneitz operator in the mathematics literature.
Higher dimensional generalizations have been constructed; see \cite{Juhl} for a review.
The above can be generalized to an arbitrary number of marginal operators and their sources
$J^I$  in such a way that the anomaly is invariant under reparametrizations of the conformal manifold whose local
coordinates are $J^I$. For explicit results in
$d=2$ and $d=4$,  see e.g. \cite{Gomis}.

Relevant operators are also easily incorporated. For instance, the source of a dimension two operator in $d=4$
transforms as $J\to e^{-2\s}J$ and
one has $\Delta_c=1$, while the source of a dimension three operator in $d=4$ transforms as $J\to e^{-\s }J$.
In this case one finds  $\Delta_c=\square-{1\over 6}R$ which transforms as $\Delta_c\to e^{-3\s}\Delta_c e^{\s}$
under Weyl rescaling of the metric.

We now turn to irrelevant operators, i.e. $\Delta>d$. From the structure of \eqref{Anomaly2ptp} we conclude that
the Weyl variation of the generating functional should contain a term \eqref{dWn} with $n>{d/2}$ and the
full expression in curved space-time should be Weyl invariant.
Assuming that the sources transform homogeneously according to their dimension, i.e.
\begin{equation}\label{deltanaive}
\delta_\s g_{\mu\nu}=2\,\s\,g_{\mu\nu}\,,\qquad\hbox{and}\qquad\delta_\s J=(\Delta-d)\,\s\,J\,,
\end{equation}
this would require the existence of an operator
\begin{equation}
\Delta_c=\square^n+\hbox{curvature terms}
\end{equation}
such that
\begin{equation}\label{Paneitz_n}
\Delta_c(e^{2\s}g)=e^{-\Delta\s}\Delta_c(g)e^{-(\Delta-d)\s}
\end{equation}
where, as before, $n=\Delta-d/2$.

The construction of Weyl anomalies for irrelevant {\it vs.} relevant and marginal operators is, however, quite different.
This is made evident by a theorem of Gover and Hirachi which excludes the existence of local operators with the
simple transformation \eqref{Paneitz_n}. More explicitly, it states
(see e.g. \cite{Juhl} for background, review and references to the original literature):

{\it On any (pseudo-)Riemannian manifold $(M,g)$ of dimension $d$ there exists
a differential operator of the form
\begin{equation}
\Delta_c=\square^n+\hbox{\rm lower order terms}
\end{equation}
such that
\begin{equation}
\Delta_c(e^{2\s}g)=e^{-({d\over2}+n)\s}\Delta_c(g)e^{+({d\over2}-n)\s}
\end{equation}
with the restriction $1\leq n\leq {d\over2}$ for even $d$ and no restriction for odd $d$.}

This operator is natural in the sense that it is constructed from the metric and the lower order terms
vanish in flat backgrounds.
Its construction is essentially holographic.
In general the operator is not unique. For instance, for $n=d/2=3$ there is a two-parameter family
\cite{Ruben}.
As there is no issue in odd dimensions, we will restrict to even $d$ in the following and,
more specifically, to $d=2$ and $d=4$ and to integer $\Delta\geq d$.

The theorem explicitly excludes the existence of such an operator for $n>d/2$.
Since a diffeomorphism invariant generating functional, whose anomaly
reduces to $J\square^n J$ in a flat metric  should exist, and the analytic structure underlying the Osborn Equation
is valid for any finite number of terms in an  expansion in the sources of irrelevant operators,
there has to be a way to avoid the conclusions of the theorem.
This obviously requires a modification in the structure of the equation.
In view of the above theorem,
the existence of beta-functions for the sources $J^i$, including those induced by the irrelevant operators,
is not sufficient to lead to a consistent equation.
The novel feature, which is only possible in the presence of sources
for irrelevant operators\footnote{Simple dimensional arguments and locality immediately imply that
a non-trivial beta-function for the metric can only occur in the presence
of sources for irrelevant operators. For marginal operators with sources $J$ the only possible modification is
$\delta g_{\mu\nu}=2\,\s\, f(J)\, g_{\mu\nu}$.},  is that the Weyl-transformation of the metric
also receives corrections. This is  a consequence of logarithms in correlation functions of the energy 
momentum tensor with irrelevant operators, in much the same way as the beta-functions for the scalar sources 
arise from logarithms in correlation functions of scalar operators.

We will now present the detailed argument for the existence of a beta-function for the metric in the presence of sources
of integer dimension irrelevant operators. To this end we
analyze the general form of the $\langle T_{\mu\nu}\,{\cal O\,O}\rangle$ three-point function.\footnote{A comprehensive
analysis of this and other CFT three-point functions in momentum space was performed in \cite{Bzowski}.
The singularity structure, which contains all the information relevant for us, can, however, be easily extracted
without knowing the full correlation function. We use the results of \cite{Petkos}, whose notation we also adopt.}
It is fixed by the
symmetries, i.e. conservation and tracelessness of $T_{\mu\nu}$,
and the normalizations of the
$\langle{\cal O\,O}\rangle$ and $\langle T_{\mu\nu}\,T_{\rho\sigma}\rangle$  two-point functions.
Comparing it
to the latter, which is also completely fixed by the symmetries except for its normalization $c_T$, one
can extract the following terms in the ${\cal OO}$ OPE
\begin{equation}\label{OO_OPE}
\begin{aligned}
{\cal O}(x)\,{\cal O}(0)&\sim {N\over x^{2\Delta}}+\dots+{a\over c_T}{1\over x^{2 \Delta+2-d}} T_{\a\b}(0)\,x^\a x^\b
+{a\over 2 c_T} {1\over x^{2\Delta+2-d}}\p_\g T_{\a\b}(0)\,x^\a x^\b x^\g\\
&\quad+{a\over 8(d+3)c_T}
{1\over x^{2\Delta+2-d}}\Big((d+4)\,\p_\g\p_\d T_{\a\b}(0)\,x^\g x^\d x^\a x^\b
-\square T_{\a\b}(0)\,x^2 x^\a x^\b\Big)+\dots\\
&\qquad +{1\over x^{2\Delta-\tilde\Delta+n}}\tilde{\cal O}_{\a_1\dots\a_n}x^{\a_1}\cdots x^{\a_n}(0)
+\dots
\end{aligned}
\end{equation}
Here
\begin{equation}
a=-{d\,\Delta N\over S_d\,(d-1)} \qquad\qquad \langle{\cal O}(x)\,{\cal O}(0)\rangle={N\over x^{2\Delta}}\qquad
S_d={\rm Vol}(S^{d-1})={2\pi^{d/2}\over\Gamma(d/2)}
\end{equation}
In $d=4$, in the normalization of the type B Weyl anomaly
as $g^{\mu\nu}\langle T_{\mu\nu}\rangle=c\,C^2$ ($C_{\mu\nu\rho\s}$ being the Weyl tensor)
one finds \cite{Erdmenger}
\begin{equation}
c_T=-{640\over\pi^2}c
\end{equation}
The terms in the last line of \eqref{OO_OPE}
are generic representatives of other primaries (and their descendants). They
will contribute to correlation functions such as $\langle T_{\mu\nu} T_{\rho\s} {\cal O\,O}\rangle$. This
will play a role in $d=2$ which we discuss in Section \ref{sectiond_2}.
Quite generally, if, suppressing possible tensor structures of the operators,
\begin{equation}
{\cal O}_j(x)\,{\cal O}_k(0)={c_{jk}{}^l\over|x|^{\Delta_j+\Delta_k-\Delta_l}}{\cal O}_l(0)+\dots\quad\hbox{with}\quad
\Delta_j+\Delta_k-\Delta_l =d+2\,n
\end{equation}
then there is a beta function for the source of ${\cal O}_l$, $\beta^l(x)\sim c_{jk}{}^l\, K^{(n)} (J^j\, J^k)(x)$
where $K^{(n)}$ is an $n$-th order differential operator which acts on the sources, therefore it vanishes
for constant sources if $n>0$.
The above discussion does not yet take into account special properties such as conservation and tracelessness of e.g.
${\cal O}_l=T_{\a\b}$. They imply that terms in the beta-function for the metric proportional
to the metric itself cannot be obtained in this way.

Using the OPE \eqref{OO_OPE} one finds (the derivatives act on $T(0)$)
\begin{equation}\label{TOO}
\begin{aligned}
\langle T_{\mu\nu}&(x){\cal O}(y){\cal O}(0)\rangle
={a\over c_T}{y^\a y^\b\over y^{2\Delta+2-d}}\langle T_{\mu\nu}(x)T_{\a\b}(0)
+{a\over 2 c_T}{y^\a y^\b y^\g\over y^{2\Delta+2-d}}\p_\g\langle T_{\mu\nu}(x)T_{\a\b}(0)\rangle\\
&+{a\over 8(d+3) c_T}{1\over y^{2\Delta+2-d}}\Big((d+4)y^\a y^\b y^\g y^\d\p_\g\p_\d-y^2 y^\a y^\b\square\Big)
\langle T_{\mu\nu}(x)T_{\a\b}(0)\rangle+\dots
\end{aligned}
\end{equation}
The first few terms on the r.h.s., their number depending on $d$ and $\Delta$, need regularization.
This introduces a scale $\mu$ and renders $\mu{d\over d\mu}$ of this
correlator non-zero. If one takes into account tracelessness and conservation of $T_{\a\b}$, one finds that the terms with
up to $n$ derivatives contribute, where
$2\Delta+2-2 d= 2(n+2)$. This is easily shown in momentum space.
E.g. for $d=4$ and $\Delta=5$, which will be analyzed in detail below,  only the first term
needs to be considered.
This leads to a beta-function for the source of $T_{\a\b}$ if $\Delta>d$, as can also be seen by simple dimensional
analysis. This also reveals that the beta-function for the metric vanishes for constant sources.

For $\Delta>d$, the three-point function
$\langle T_{\mu\nu}{\cal OO}\rangle$ has a double log in even dimensions $>2$. This follows from the second term in
\eqref{OO_OPE}. For $\Delta=d$ the coefficient of $T_{\a\b}$ is singular and regularization introduces
 a scale $\mu$.
The derivative w.r.t.
this scale vanishes because of tracelessness on $T$. For $\Delta>d$ it is proportional to
$\p^\a\p^\b \square^{\Delta-d-1}\delta^{(d)}(y)$.
In even $d>2$ the two-point function $\langle T_{\mu\nu}T_{\a\b}\rangle$ provides an additional log.
In $d=2$ this two-point function does not have a log but a $1/p^2$ singularity in momentum space.
One can explicitly verify the appearance of $(\log\Lambda^2)^2$ terms in $\langle T_{\mu\nu}\,\phi^n\,\phi^n\rangle$
for the free scalar in $d=4$  if $n\geq 5$.

On the l.h.s. of the Osborn Equation the logarithms will be cancelled exactly due to the presence
of the metric beta-function. The general structure is therefore that in correlators of irrelevant operators with
energy-momentum tensors, multiple logarithms appear and this necessitates the existence of a metric beta-function
such that the l.h.s. of the equation is a local expression, the Weyl anomaly.
We propose therefore that in the presence of irrelevant operators with sources $\bar J$ the operator of
\eqref{deltaW} is modified to:
\begin{equation}\label{ddeltaW}
\delta_\s W=\int\,d^d x\Big\{ \beta^g_{\mu\nu}(\bar J ,g_{\mu\nu};\s) {\d\over\d g_{\mu\nu}(x)}
+\sum_i\beta_i\big(\{J_j\},g_{\mu\nu};\s \big){\d\over\d J_i(x)}\Big\}W
\end{equation}
where $\beta^g_{\mu\nu}$  is the metric beta-function. It  contains, in addition to the dimensional term $2\,\s(x)g_{\mu\nu}(x)$,
contributions which involve the sources for the irrelevant operators. The second sum is over all scalar sources.

We may also consider $\langle T_{\mu\nu}(x_1)T_{\rho\s}(x_2){\cal O}(x_3)\rangle$. Again, tracelessness and
conservation fixes it up to one overall constant \cite{Petkos}.
In $d=2$ these conditions imply that it vanishes except if $\Delta=0$ which, in a unitary theory, means
${\cal O}={\mathds 1}$ or if $\Delta=4$, which means ${\cal O}=T^{\mu\nu}T_{\mu\nu}$.
Expressed in a chiral basis this can be recognized as Zamolodchikov's $T\bar T$\cite{Zamo}.
For general $d$, the singularity
at $x_1\to x_2$ is $x_{12}^{\Delta-2 d}$, which requires regularization for even
$\Delta>d$. On the other hand, when in addition also $x_2\to x_3$, the singularity  is $x_{23}^{-2\Delta}$.
This leads to an anomaly linear in the source and quadratic in $h_{\mu\nu}$, where
$g_{\mu\nu}=\eta_{\mu\nu}+ h_{\mu\nu}$.

The question of energy-momentum correlators  which contain multiple logarithms is a very interesting one and in
general depends on the specific structure of the CFT. In the present paper we will mainly discuss the multiple
logarithms generated by the interplay of the $\langle \cal O \cal O\rangle$, $\langle T \cal O \cal O\rangle$ and
$\langle T T \rangle $ correlators. They exist in any CFT once the integer dimensional irrelevant operator $\cal O$ is present.

\section{$\mathbf{d=4}$}

In this section we make the above discussion concrete and derive a non-zero beta function
for the metric by combining the regularized correlation function  $\langle T_{\mu\nu}{\cal O\,O}\rangle$ with
the Osborn Equation.\footnote{This extends the analysis of \cite{vanRees} where only scalar operators were considered and
the extension to a curved background metric was not addressed.} A concrete realization, on which
the general arguments can be explicitly checked, is a free scalar field
$\phi$ with $\phi^n$ being an operator of dimension $\Delta=n$.

We start by including the lowest integer dimension irrelevant operator, which has dimension $\Delta=5$.
We denote its source by $\rho$. Some terms in the generating functional, which will play a role in the following discussion, are
\begin{equation}\label{Wpert}
\begin{aligned}
W[\rho,h]&=
{1\over 2}\int d^4x\, d^4y\langle {\cal   O}(x)\,{\cal O}(y)\rangle\rho(x)\,\rho(y)+
{1\over2}\int d^4 x\, d^4 y \langle T_{\mu\nu}(x)\,T_{\a\b}(y)\rangle\,h^{\mu\nu}(x)\,h^{\a\b}(y)\\
&\qquad+{1\over2}\int d^4 x\, d^4 y\, d^4 z \langle T_{\mu\nu}(x)\,{\cal O}(y)\,{\cal O}(z)\rangle \,h^{\mu\nu}(x)\,\rho(y)\,\rho(z)+\dots
\end{aligned}
\end{equation}
All correlation functions are computed in the original CFT. These terms are present in any CFT which has a dimension
five operator. The first and third terms in \eqref{Wpert} should be considered  together since they are
related by the diffeomorphism Ward identity which we will always implement.

For $d=4$ and $\Delta=5$  \eqref{TOO} becomes
(the derivatives are w.r.t. the argument of $T_{\a\b}$)
\begin{equation}\label{TOOd=4}
\begin{aligned}
\langle T_{\mu\nu}(x)\,{\cal O}(y)\,{\cal O}(0)\rangle
&={a\over c_T}{y^\a y^\b\over y^8}\langle T_{\mu\nu}(x)T_{\a\b}(0)
+{a\over 2 c_T}{y^\a y^\b y^\g\over y^8}\p_\g\langle T_{\mu\nu}(x)T_{\a\b}(0)\rangle\\
\noalign{\vskip.2cm}
&+{a\over 56\, c_T}\Big(8{y^\a y^\b y^\g y^\d\over y^8}\p_\g\p_\d-{y^\a y^\b\over y^6}\square\Big)
\langle T_{\mu\nu}(x)T_{\a\b}(0)\rangle+\dots
\end{aligned}
\end{equation}
We have only displayed those terms which require regularization, which introduces a renormalization scale and
hence type B anomalies and beta-functions.
Due to tracelessness and conservation of $T_{\a\b}$ only the first term contributes and one finds
\begin{equation}
\mu{d\over d\mu}\langle  T_{\mu\nu}(x){\cal O}(y){\cal O}(0)\rangle={a\,\pi^2\over 24\,c_T}\p^\a\p^\b\d^{(4)}(y)
\langle T_{\mu\nu}(x)T_{\a\b}(0)\rangle
\end{equation}
This can be derived either in position
space using differential regularization \cite{Freedman} or, more straightforwardly,
in momentum space. Therefore
\begin{equation}
\begin{aligned}
\mu{d\over d\mu}W
&={a\,\pi^2\over24\,c_T}\int d^4 x d^4 z\,\langle T_{\mu\nu}(x)\, T_{\a\b}(z)\rangle h^{\mu\u}(x)\,\rho(z)\p^\a\p^\b\rho(z)+\dots\\
&={a\,\pi^2\over24\,c_T}\int d^4 z\,\rho(z)\,\p_\a\p_\b\rho(z)\,{\d\over\d h^{\a\b}(z)}W+\dots
\end{aligned}
\end{equation}
We require that $W$ satisfies the Osborn Equation \eqref{Osborneq},  i.e. after acting with the Weyl variation
operator the terms with single logarithms are cancelled. This cancellation is between a contribution
of the third  term (a variation of a double logarithm)  and a contribution of the second term which needs
a ${\cal O}(\rho^2)$ contribution to the Weyl variation of the metric
\begin{equation}\label{dqg}
\delta_\s g_{\mu\nu}=2\,\s\, g_{\mu\nu}+{a\,\pi^2\over 24\, c_T}\s\,\rho\,\p_\mu\p_\nu\rho+\dots
\end{equation}
Here $\dots$ are possible terms contributing to $\beta^g_{\mu\nu}$ that cannot be derived by the above analysis, which
is based on the two- and three-point functions. These are terms which
either vanish in flat space or for constant $\s$, terms which are proportional to $g_{\mu\nu}$ and, of course, terms
of higher order in $\rho$.
Once the logarithms in the variation are cancelled, there is a local anomaly which to this order -- second order in $\rho$
and zeroth order in the metric -- comes from the first term in \eqref{Wpert} and it is $\sim\rho\,\Box^3\rho$.

We will now determine all terms to ${\cal O}(\rho^2)$,  but to all orders in the metric,
seeded by the lowest terms \eqref{Wpert} by performing a cohomological analysis.
Once this is achieved, we determine the anomaly by imposing WZ consistency.

To find the complete beta-function for the metric to ${\cal O}(\rho^2)$ or, equivalently, to second order
in derivatives, we start with the most general Ansatz for $\delta_\s g_{\mu\nu}$ to this order, which is consistent with
dimensional analysis.
We then extend and perform the cohomological analysis outlined in the Introduction for $\beta^g_{\mu\nu}(\rho,g,\s)$,
i.e. we impose  $[\d_{\s_1},\d_{\s_2}]g_{\mu\nu}=0$ which leads to \eqref{WZbeta} for $\beta^g_{\mu\nu}$.
This is solved modulo trivial solutions, i.e. local expressions $\gamma_{\mu\nu}$ s.t.
such that $\delta_\s(g_{\mu\nu}+\gamma_{\mu\nu})=2\,\s\,(g_{\mu\nu}+\gamma_{\mu\nu})$.

For the transformation of $\rho$ we have to this order
\begin{equation}\label{deltarho}
\delta_\s\rho=\s\,\rho
\end{equation}
For the metric the general Ansatz  is
\begin{equation}\label{Ansatz}
\begin{aligned}
\delta_\s g_{\mu\nu}&=2\,\s\,g_{\mu\nu}+c_1\,\s\,\nabla_\mu\rho\,\nabla_\nu\rho+c_2\,\rho^2\,\nabla_\mu\nabla_\nu\s
+c_3\,\s\,\rho\,\nabla_\mu\nabla_\nu\rho+c_4\,\rho(\nabla_\mu\rho\,\nabla_\nu\s+\nabla_\nu\rho\,\nabla_\mu\s)\\
\noalign{\vskip.2cm}
&+c_5\,\s\,g_{\mu\nu}\,\rho\,\square\rho+c_6\,\s\,g_{\mu\nu}\,\nabla^\a\rho\,\nabla_\a\rho
+c_7\,\s\,g_{\mu\nu}\,\rho^2\,R+c_8\,g_{\mu\nu}\,\rho^2\,\square\s\\
\noalign{\vskip.2cm}
&+c_9\,g_{\mu\nu}\,\rho\,\nabla^\a\rho\,\nabla_\a\s
+c_{10}\,\s\,R_{\mu\nu}\,\rho^2
\end{aligned}
\end{equation}
Imposing consistency \eqref{WZbeta}  leaves, in general dimension, six solutions:  the ones parametrized by
$c_2, c_4, c_8$ and $c_9$, which can be shown to be cohomologically trivial in the sense described above;
\begin{equation}\label{solWZmetric}
\begin{aligned}
\s\left(R_{\mu\nu}\,\rho^2+(d-2)\rho\,\nabla_\mu\nabla_\nu\rho-(d-1)g_{\mu\nu}(\nabla\rho)^2
+g_{\mu\nu}\,\rho\,\square\rho\right)&\equiv\s\,\rho^2\,\hat R_{\mu\nu}\\
\end{aligned}
\end{equation}
and
\begin{equation}\label{sigtilde}
g_{\mu\nu}\,\s\left(R\,\rho^2+2(d-1)\rho\,\square\rho-d(d-1)(\nabla\rho)^2\right)
\equiv\s\,\rho^2\,\hat g_{\mu\nu}\,\,\hat R
\end{equation}
Here $\hat R$ is the curvature computed with the Weyl-invariant metric
$\hat g_{\mu\nu}={1\over\rho^2}g_{\mu\nu}$.
We choose the first solution in \eqref{solWZmetric}. It contains the ``seed" \eqref{dqg} and extends it
to all orders in $g_{\mu\nu}$.  Adding \eqref{sigtilde} would not modify the result which we are going to find for the anomaly.
A general discussion of beta-functions with the structure of the second solution,
i.e. non-trivial consistent solution proportional to $g_{\mu\nu}$, multiplied by a Weyl invariant structure, will be
given later.

We now make the general Ansatz for the anomaly ${\cal A}(\rho,g_{\mu\nu})$
to ${\cal O}(\rho^2)$ or, equivalently, to sixth order in derivatives,
and require that it satisfies the WZ consistency condition \eqref{WZanomaly}.
We use the normalization
\begin{equation}\label{normdg}
\delta_\s g_{\mu\nu}=2\,\s\,g_{\mu\nu}+\a\,\s\,\rho^2\,\hat R_{\mu\nu}
\end{equation}
where $\a$ is related to $a$ in \eqref{TOO}, i.e. to the normalization of $\langle T_{\mu\nu}\,{\cal O\,O}\rangle$,
as will be made explicit below.
One then finds the following solution of the WZ consistency condition
\begin{align}\label{anomalyd_4}
{\cal A}=&c\,C_{\mu\nu\rho\s}C^{\mu\nu\rho\s}+\a\,c\Bigg\{\square\rho\,\square^2\rho
-{13\over8}R\,R^{\mu\nu}R_{\mu\nu}\,\rho^2+{53\over 162}R^3\,\rho^2
+{4\over3}R^{\mu\nu}R^{\rho\s}R_{\mu\rho\nu\s}\,\rho^2\nonumber\\
\noalign{\vskip.0cm}
&-{1\over8}R\,R_{\mu\nu\rho\s}R^{\mu\nu\rho\s}\rho^2
+{43\over72}R_{\mu\nu\rho\s}R^{\mu\nu\a\b}R_{\a\b}{}^{\rho\s}\,\rho^2-{35\over72}R^2\,\rho\,\square\rho
+{25\over24}R_{\mu\nu\rho\s}R^{\mu\nu\rho\s}\rho\,\square\rho\nonumber\\
\noalign{\vskip.2cm}
&-{1\over36}\nabla^\mu R\,\nabla_\mu R\,\rho^2+{167\over12}R^{\mu\nu}R_{\mu\nu}\,\nabla^\a\rho\,\nabla_\a\rho
-{101\over24}R^2\,\nabla^\a\rho\,\nabla_\a\rho\nonumber\\
\noalign{\vskip.2cm}
&-{79\over24}R^{\mu\nu\rho\s}R_{\mu\nu\rho\s}\,\nabla^\a\rho\,\nabla_\a\rho-{1\over3}R\,\square\nabla^\mu\rho\,\nabla_\mu\rho
-{10\over9}R^{\mu\nu}\,\nabla_\mu\nabla_\nu R\,\rho^2+{7\over9}R^{\mu\nu}\,R\,\rho\,\nabla_\mu\nabla_\nu\rho\nonumber\\
\noalign{\vskip.2cm}
&+{1\over36}\square R\,\rho\,\square\rho-{16\over9}R\,(\square\rho)^2+\nabla^\mu R\,\nabla_\mu\rho\,\square\rho
+{1\over6}R\,\rho\,\square^2\rho
-4\,R^{\mu\nu}\,\nabla_\mu\rho\,\square\nabla_\nu\rho\\
\noalign{\vskip.2cm}
&-{37\over18}R_{\mu\nu}\nabla^\mu R\,\rho\,\nabla_\nu\rho-22\,R_{\mu}{}^\a R_{\nu\a}\,\nabla^\mu\rho\,\nabla^\nu\rho+{116\over9}R^{\mu\nu}\,R\,\nabla^\mu\rho\,\nabla^\nu\rho\nonumber\\
\noalign{\vskip.2cm}
&-13\, R^{\a\b}R_{\mu\a\nu\b}\,\nabla^\mu\rho\,\nabla^\nu\rho-{5\over18}\nabla^\mu\nabla^\nu R\,\rho\,\nabla_\mu\nabla_\nu\rho
-{5\over9}R\,\nabla_\mu\nabla_\nu\rho\,\nabla^\mu\nabla^\nu\rho\nonumber\\
\noalign{\vskip.2cm}
&-5\, R^{\b\g}\,\nabla_\g R_{\a\b}\,\rho\,\nabla^\a\rho
-{8\over3}R_{\a}{}^\g R^{\a\b}\,\rho\,\nabla_\b\nabla_\g\rho
+{10\over3}R^{\b\g}\,\nabla^\a\rho\,\nabla_\g\nabla_\b\nabla_\a\rho\nonumber\\
\noalign{\vskip.2cm}
&+{5\over6}\square R^{\mu\nu}\,\rho\,\nabla_\mu\nabla_\nu\rho
+{22\over3}R^{\mu\nu}\,\nabla_\mu\nabla_\nu\rho\,\square\rho
-{5\over3}\nabla^\mu R^{\a\b}\,\nabla_\mu R_{\a\b}\,\rho^2
\Bigg\}+{\cal O}(\rho^4)\nonumber
\end{align}
This anomaly is entirely type B, as it does not vanish for constant Weyl parameter $\s$, i.e.
neither the ${\cal O}(\rho^0)$ nor the ${\cal O}(\rho^2)$ parts of
${\cal A}$ are total derivatives.
The second term is what we obtained before to lowest order. The remaining terms are higher order in the metric
and they vanish in flat space.
They implement  diffeomorphism invariance.
Note the interplay between the $\rho$-independent ordinary type B Weyl anomaly and
the $\rho$-dependent terms, which is possible due to the $\rho$-dependence of $\delta_\s g_{\mu\nu}$, which
is in turn required to avoid the conclusions from the theorem of Gover and Hirachi.

Comparing \eqref{normdg} with \eqref{dqg} we identify
\begin{equation}
2\,\a={a\,\pi^2\over 24\,c_T}={N\,\pi^2\over 3^2\cdot 2^9\,c}
\end{equation}
The anomaly in the two-point function $\langle{\cal OO}\rangle$
derived from the above anomaly polynomial is
\begin{equation}
2\,c\,\a\,\square^3\d^{(4)}(x)={N\,\pi^2\over 2^9\cdot 3^2}\square^3\delta^{(4)}(x)
\end{equation}
which agrees with the (anomalous) change of \eqref{OO} under scale transformation.

By adding a cohomologically trivial term
\begin{equation}
\Delta W={\a\,c\over 24}\int d^4 x\sqrt{g}\Big(6\,\square\rho\,\square^2\rho-2\,R\,\rho\,\square^2\rho-R\,(\square\rho)^2\Big)
\end{equation}
one can eliminate the second
term in \eqref{anomalyd_4} in favour of $\rho\,\square^3\rho$; of course the curvature
dependent terms will also change.

The anomaly \eqref{anomalyd_4} can be understood as ``seeded" by the correlator of two irrelevant operators.
Once the metric beta-function is turned on, the usual forms of the Weyl anomalies (type A and B) are no longer
valid since the Weyl variation of the metric has changed. As a consequence the corresponding expressions will be deformed
by the metric beta-function. In this sense \eqref{anomalyd_4} can be understood also as  the appropriate
deformation of the $"c"$ Weyl anomaly, as the presence of the first term shows.
The  information contained in \eqref{anomalyd_4} is universal, i.e. independent on the regularization scheme.
It is theory dependent through the  two parameters which give the coefficients of single logarithms
in the correlator of two energy-momentum tensors and two irrelevant operators, respectively.

A similar deformation occurs for the type A Euler anomaly in $d=4$.
The cohomologically nontrivial anomaly starting with the Euler density, when the metric variation
contains the deformation \eqref{normdg}, is
\begin{equation}\label{Eulerseed}
\begin{aligned}
{\cal A}=&a\,E_4 + \a\,a\Bigg\{{28\over135}R^3\,\rho^2-{6\over5}\,R^{\mu\nu}\,R^{\rho\s}\,R_{\mu\rho\nu\s}\,\rho^2
-{7\over10}\,R\,R^{\mu\nu\rho\s}\,R_{\mu\nu\rho\s}\,\rho^2\\
&+{14\over 15}\,R_{\mu\nu\rho\s}R^{\rho\s\a\b}R_{\a\b}{}^{\mu\nu}\rho^2
-{1\over15}R\,\square R\,\rho^2+{2\over3}R^2\,\rho\,\square\rho+{58\over5}R^{\mu\nu}R_{\mu\nu}(\nabla\rho)^2\\
\noalign{\vskip.2cm}
&-{64\over15}\,R^2\,(\nabla\rho)^2+{1\over5}R^{\m\nu\rho\s}R_{\mu\nu\rho\s}\,(\nabla\rho)^2
-12\,R_{\mu\rho}R_\nu{}^\rho\,\nabla^\mu\rho\,\nabla^\nu\rho+8\,R\, R_{\mu\nu}\,\nabla^\mu\rho\,\nabla^\nu\rho\\
\noalign{\vskip.2cm}
&-8\,R^{\rho\s}R_{\mu\rho\nu\s}\,\nabla^\mu\rho\,\nabla^\nu\rho
+{2\over5}\,R^{\mu\nu}\,\square R_{\mu\nu}\,\rho^2-2\,R^{\mu\nu}R_{\mu\nu}\,\rho\,\square\rho
+{7\over5}\,\nabla^\mu R^{\nu\rho}\,\nabla_\nu R_{\mu\rho}\,\rho^2\\
&-{7\over10}\,\nabla^\rho R^{\mu\nu}\,\nabla_\rho R_{\mu\nu}\,\rho^2-{4\over5}
\nabla^\s\nabla^\rho R^{\mu\nu}\,R_{\mu\rho\nu\s}\,\rho^2
-{7\over20}\,\nabla^\a R^{\mu\nu\rho\s}\,\nabla_\a R_{\mu\nu\rho\s}\,\rho^2\Bigg\}
\end{aligned}
\end{equation}
Here
\begin{equation}
E_4= R^2-4\, R^{\mu\nu}R_{\mu\nu}+R^{\mu\nu\rho\s}R_{\mu\nu\rho\s}
\end{equation}
is the Euler density normalized s.t.  $\int_{S^4}\sqrt{g}\,E_4=64\,\pi^2$.
This deformed anomaly does not introduce any new universal parameter.

The terms in the Osborn Equation, whose variation is given by \eqref{Eulerseed}, have joint (single) logarithmic  and
powerlike ("type A") singularities.
They can be understood as originating in a decomposition of correlators of two ${\cal O}$ operators with three
energy-momentum tensors, where the two irrelevant operators connect through the energy-momentum tensor conformal block.

It might be interesting to note that while \eqref{anomalyd_4} vanishes in a maximally symmetric space, e.g. $S^4$,
for constant source $\rho$, this is not the case for \eqref{Eulerseed}. On the other hand \eqref{Eulerseed} vanishes
in flat space while \eqref{anomalyd_4} does not (for non-constant sources).

The deformation of the cohomologically trivial solution $\int d^4x\sqrt{g}\, \s\, \Box R$ of the Wess-Zumino condition continues to
be trivial, being the variation of the local term $\int d^4 x\sqrt{g}\,R^2$ under the deformed operator.

To the anomalies determined above, related to  the order $\rho^2 $ contribution in the metric beta-function,
we can add ``ordinary" anomalies, i.e. type B anomalies with quadratic $\rho$ dependence.
If the new  structures lead to Weyl invariant anomaly integrands  when $g_{\mu\nu}$ transforms naively,
these anomalies are also consistent to order $\rho^2$ when the metric beta-function is taken into account.
There are eight such structures.
Two of them are simple as they are constructed from the Weyl invariants
\footnote{There is a second
contraction of three Weyl tensors, but in dimensions $<6$ it is not independent of $I_1$.}
\begin{equation}\label{I12}
\begin{aligned}
I_1&=C_{\mu\nu\rho\s}C^{\rho\s\a\b}C_{\a\b}{}^{\mu\nu}\\
\noalign{\vskip.2cm}
I_2&=-\nabla^\a R^{\mu\nu\rho\s}\,\nabla_\a R_{\mu\nu\rho\s}
-16\, R_{\mu\rho\nu\s}\,\nabla^\s\nabla^\rho R^{\mu\nu}
-10\nabla^\mu R^{\nu\rho}\,\nabla_\mu R_{\nu\rho}
+12 \, \nabla^\mu R^{\rho\s}\,\nabla_\r R_{\mu\s}\\
&\qquad+8\, R^{\mu\nu}\,\square R_{\mu\nu}+{2\over3}\,\nabla^\mu R\,\nabla_\mu R
-{4\over3}R\,\square R-{2\over3}R\, R^{\mu\nu\rho\s}R_{\mu\nu\rho\s}+8\, R^{\mu\nu}R^{\rho\s}R_{\mu\rho\nu\s}\\
&\qquad\quad-8\, R_\mu{}^\nu R_\nu{}^\rho R_\rho{}^\mu+{4\over3} R\, R^{\mu\nu}R_{\mu\nu}
-{2\over9}R^3
\end{aligned}
\end{equation}
There is a third simple solution which involves derivatives of $\rho$:
\begin{equation}\label{I3}
\begin{aligned}
I_3&= {\hat R}\, C_{\mu\nu\a\b} C^{\mu\nu\a\b}
\end{aligned}
\end{equation}
We will not write down the other five solutions which, as $I_3$, also contain
terms which vanish for constant source $\rho$.
$I_i$, $i=1,2$,  transform under Weyl rescaling of the metric as $\delta_\s I_i=-6\,\s\,I_i$ while $I_i,\, i=3,\dots,8$, transform
under a joint rescaling of the metric and $\rho$ as $\delta_\s I_i=-4\,\s\,I_i$. They do not exist in the absence of $\rho$.

The type B anomalies constructed from these structures are
$\int d^4x\, \s \sqrt{g}\,\rho^2 I_i$ for $i=1,2$
and $\int d^4 x\, \s \sqrt{g}\, I_i,\,i=3,\dots,8,$ respectively.
They correspond to variations of single logarithms in the generating functional. As we will see in the
continuation, their appearance could also be related to sources for higher dimension operators.
Their coefficients are theory dependent and in particular can vanish in
counterdistinction from the anomaly \eqref{anomalyd_4}.

We remark that the anomaly based on $I_3$ above can be understood as a deformation of the $``c"$ trace anomaly
by  \eqref{sigtilde}. It is then natural to conjecture that contributions of the type \eqref{sigtilde} to the metric
beta-function proportional to $g_{\mu\nu}$ do not represent genuine modifications required by the appearance of multiple
logarithms, but simply reproduce ordinary higher order type B anomalies.

We should discuss the stability of these structures when beta-functions for other operators induced
by the dimension five operator are also included. If we include only terms up to quadratic order in $\rho$, the only
other operators which have to be considered are relevant, marginal and dimension six irrelevant operators.
In order that the induced beta-functions
contribute in the Osborn Equation
we should limit the new couplings to first order.
We can then invert the logic and start with contributions of the new operators at linear order in their sources
and subsequently study the deformations to order $\rho^2$.

We begin by including  the source $\tau$ for an operator
$\tilde {\cal O}$ of dimension six. Generically $\langle{\cal OO\tilde O}\rangle\neq 0$,
e.g. for a free scalar field with ${\cal O}=\phi^5$ and $\tilde{\cal O}=\phi^6$.
As a consequence there is a beta-function for $\tau$ given by
\begin{equation}\label{deltatau}
\delta_\s\tau=2\,\s\,\tau+\b\,\s\,\rho^2
\end{equation}
Here $\b$ is proportional to the OPE coefficient $c_{\cal OO\tilde O}$. Therefore the dimension six operator can
indeed give contributions to the Osborn Equation to order $\rho^2$ and we should study its influence to linear order in $\tau$.

Repeating the cohomological analysis for possible metric beta-functions, one finds that there is no
cohomologically nontrivial solution to order $\tau$. As a consequence when the deformation \eqref{deltatau}
is included there will be no modification of \eqref{solWZmetric} due to the presence of $\tau$. We can analyze
now the Weyl anomalies to order $\tau$.
The correlator $\langle T T \tilde{\cal O}\rangle$ has a logarithmic singularity and we expect therefore that type B
anomalies linear in $\tau$ will exist which are at least quadratic in the curvature.
Indeed there are four such solutions of the WZ-condition. Clearly $\tau\,I_i$ for $i=1,2$ are solutions.
The other two can e.g. be obtained
if one replaces $\rho=\sqrt{\tau}$ and requires regularity in $\tau$. This is satisfied by two of the $I_i$, $i=4,\dots,8$,
say $I_4$ and $I_5$:
\begin{equation}\label{tauanomaly}
\delta_\s W=\int d^4 x\,\sqrt{g}\,\s\,\Big\{\tau\big( c_1\, I_1(g)+c_2\,I_2(g)\big)+c_4\,I_4(g,\sqrt{\tau})
+c_5\,I_5(g,\sqrt{\tau})\Big\}
\end{equation}
with $I_{1,2}$ as in \eqref{I12}.
In the presence of the $\rho$-dependent terms in \eqref{deltatau}, the anomaly \eqref{tauanomaly} is no longer a consistent
solution of the Osborn Equation. The deformation of the anomaly to ${\cal O}(\rho^2)$, ignoring terms of order
$\tau \rho^2$,  are precisely the
$\int d^4x\sqrt{g}\,\s \rho^2 I_i(g),  i=1,2$ and $\int d^4x\sqrt{g}\,\s I_i(g,\rho),\,i=4,5$
discussed before. These anomalies do not mix
with \eqref{anomalyd_4} and now they acquire a natural interpretation: they originate in the coupling of two
irrelevant dimension five operators to the dimension six operator conformal block in  the correlator with
energy-momentum tensors. Therefore the dimension six operator, even though it has
an induced beta-function to order $\rho^2$, does not change the structure of the metric  beta-function and
Weyl anomaly produced by the dimension five operator alone.

We can perform a similar analysis for a marginal (but not necessarily exactly marginal)
operator $M$ coupled to a source $J$.
We analyze  the cohomology problems  linear in $J$.
The induced beta-function for $J$ is
\begin{equation}\label{indJ}
\delta_{\s} J = \gamma\,\s\, {\hat R}
\end{equation}
The coefficient $\gamma$ is proportional to the $c_{M\cal OO}$ OPE coefficient. This obviously satisfies
the consistency condition and is non-trivial. The appearance of $\hat R$, or rather its flat space limit, can be verified
by studying the $\langle M{\cal OO}\rangle$ correlation functions.
As follows from our general analysis,  a marginal operator can generate in the metric
beta-function only a term proportional to the metric itself with a $J$-dependent coefficient function.
As argued before such a term  cannot contribute to the genuinely new anomaly structure related to multiple logarithms
and therefore will not interfere with \eqref{solWZmetric}.
There is a type B anomaly linear in $J$ related to the standard $``c"$ Weyl anomaly
\begin{equation}\label{Jano}
\int d^4 x\,\sqrt{g}\,\s\, J \, C_{\mu\nu\a\b}\,C^{\mu\nu\a\b}
\end{equation}
After having understood the structure at linear order in $J$,
we can deform it using \eqref{indJ}. The metric beta-function
\eqref{solWZmetric} will not be changed while the deformation
of \eqref{Jano} leads to the anomaly based on the $I_3$-structure above.
We have again a new interpretation of this order $\rho^2$ anomaly as resulting from a
decomposition of the  correlator of two irrelevant operators and energy-momentum tensors, where the
irrelevant operators couple through the marginal ($J$) conformal block.
We therefore conclude that at ${\cal O}(\rho^2)$ there is no change
in  the metric beta-function \eqref{solWZmetric} or the Weyl anomaly \eqref{anomalyd_4}.
We do not include a similar
analysis for relevant operators. They also receive an induced beta-function at order $\rho^2$,
the conclusion being the same, i.e. stability of genuine $\rho^2$ contribution.

In $d=4$ we have therefore a clear distinction between contributions to the beta-function and  anomalies related to
multiple logarithms and usual type B  contributions in correlators with energy-momentum tensors. We can even
characterize the difference by singling out the conformal block through which the irrelevant operators are connected to
the energy-momentum tensors in a given decomposition scheme. It is an interesting question how this information
constraints decompositions related by crossing.

In $d=2$ the stability analysis of the metric  beta-function related to induced couplings
is much more subtle. This will be discussed in detail the next section.

Let us also stress once more that throughout we worked to order six in derivatives
or, equivalently, to order two in $\rho$.
There will be contributions at all higher higher orders in $\rho$, but they will only start contributing in
correlation functions with more than two insertions of  ${\cal O}$.
At higher orders we will also have to take into account the mixing with operators of dimension $\Delta>5$,
namely of those operators which appear in the ${\cal OO}$ OPE.

We treated in detail above the structure induced in  second order by the minimal integer dimension irrelevant operator.
This generalizes to higher dimensional operators but with increasingly  complex expressions.
We limit ourselves to a discussion of the salient features of the generalization.
We consider the anomaly seeded by the quadratic correlator of a dimension six operator whose coupling we
continue to denote by $\tau$, but working now to ${\cal O}(\tau^2)$.
We check first  whether there is a non-trivial
beta function for the metric. Making the general Ansatz for $\delta g_{\mu\nu}$ at ${\cal O}(\tau^2,\nabla^6)$,
imposing non-triviality and the WZ condition, one finds a five-parameter family of solutions.
None of the solutions contains derivatives of $\s$. This means that there is no interference between the
${\cal O}(\tau^0)$ and the ${\cal O}(\tau^2)$ terms and they satisfy the WZ condition independently.
Two of the solutions are proportional to $g_{\mu\nu}$ and, as argued above, will not
represent ``genuine" contributions related to multiple logarithms.

Among the three remaining solutions only one survives in the flat limit and can in principle be extracted from
$\langle T{\cal OO}\rangle$. In contrast to $\Delta=5$, for $\Delta=6$ more than just the first
term on the r.h.s. of \eqref{OO}  which contributes to $\langle T{\cal OO}\rangle$ needs to be considered.
More precisely, as follows from the discussion below eq.\eqref{TOO},
the second term on the r.h.s. of this equation now also contributes.

One wonders whether these solutions can be constructed, as in case $\Delta=5$ above, from
expression constructed from the invariant metric $\hat g_{\mu\nu}={1\over\tau}g_{\mu\nu}$.
For instance $\s\,\tau\,\hat R\,\hat R_{\mu\nu}$ transforms homogeneously with weight two, but it is singular
in the limit $\tau\to0$. There are nine such terms with four derivatives of which five
are regular as $\tau\to 0$. They agree with the non-trivial solutions of the WZ condition.

\section{$\mathbf{ d=2}$}\label{sectiond_2}
The basic structure in $d=2$ is the same as in higher even dimensional theories, i.e. there is a beta-function
originating in a logarithmically singular coefficient  in front of $T$ in the OPE  of two irrelevant operators $\cal O$.
But there are specific $d=2$ features which change the analysis. In particular, in $d=2$  the energy-momentum tensor
two-point function has a "type A" anomaly, i.e. the logarithmic dependence is replaced by a singular $1/\Box$ dependence.
This will be a general feature, i.e. the multiple logarithms in the various correlators studied will change
into one logarithm multiplied by integer power singularities.
More generally, even though in the diffeomorphism preserving scheme we are using
only the $SO(2,2)$ symmetry is explicit, there is the underlying Virasoro symmetry which  implies additional restrictions.
We will study again the minimal dimension irrelevant operator which in $d=2$ is $\Delta=3$, its source $\rho$ having the
Weyl transformation \eqref{deltarho}.

A concrete realization of such a CFT is a theory of three free Majorana fermions. The dimension three irrelevant operator is
\begin{equation}\label{Majorana}
{\cal O}\equiv \bar\psi_{1}\bar\psi_{2}\bar\psi_{3} \psi_{1}\psi_{2}\psi_{3}
\end{equation}
where $\bar\psi , \psi$ are the left and right Majorana fields,  respectively .

We will start by solving the cohomological problems in the most general setup to  quadratic order in $\rho$.
At this order there is an additional ``induced" source $\tau$
coupled to a dimension 4 operator $\tilde O$.
One can then check that the following transformations are non-trivial solutions of the WZ condition:
\begin{equation}\label{dg2}
\begin{aligned}
\delta_\s\rho&=\s\,\rho\\
\noalign{\vskip.1cm}
\delta_\s\tau&=2\,\s\,\tau+\a\,\s\,\rho^2\\
\noalign{\vskip-.2cm}
\delta_\s g_{\mu\nu}&=2\,\s\,g_{\mu\nu}+\tau\,\nabla_\mu\nabla_\nu\s+\a\,\s\Big(\rho\,\nabla_\mu\nabla_\nu\rho
-{1\over2}g_{\mu\nu}(\nabla\rho)^2\Big)
\end{aligned}
\end{equation}
The second term in $\delta_\s\tau$ is the ``induced" coupling whose normalization $\alpha$ reflects the
$c_{\cal O \cal O \tilde O}$ structure constant.
The coefficient of the second term in $\delta_\s g_{\mu\nu}$
has been set to one by a suitable rescaling of $\t$, but it cannot be set to zero. Indeed, in
contrast to $d=4$, the metric beta-function now depends on $\tau$ and therefore the mixing between
the dimension three and four operators cannot be neglected in the anomaly analysis.
The peculiar features of these solutions will be discussed later.

There are contributions to order $\rho^2$ also to the marginal and relevant operators but they do not contribute to the
anomaly: we will analyse later  in this Section their detailed structure and especially the differences compared with the
role of $\tau$.

The presence of the metric beta-function once again means that there is an interplay between
the purely gravitational part of the Type A anomaly, the Polyakov term $c\,R$, and the part which
depends on the sources $\rho$ and $\tau$.
The most general Ansatz for the anomaly, after subtracting variations of local terms, can be reduced to
\begin{equation}
\begin{aligned}
\delta_\s W&=\int\sqrt{g}\,\s\Big(c\,R+a_1\,R^2\,\rho^2+a_2\,\rho^2\,\square R+a_3\,R\,\rho\,\square\rho
+a_4\,R\,(\nabla\rho)^2\\
&\qquad\qquad\qquad+a_5\,\rho\,\square^2\rho+a_6\,R^2\,\tau+a_7\,\tau\,\square R+a_8\,R\,\square\tau\Big)
\end{aligned}
\end{equation}
Imposing WZ on the anomaly leads to the following unique solution
\begin{equation}
\begin{aligned}
\delta_\s W=\int\sqrt{g}\,\s\,c\Big(R+{1\over16}\a\,R^2\,\rho^2
+{1\over4}\a\,R\,\rho\,\square\rho-{1\over2}\a\,R\,(\nabla\rho)^2+{1\over4}\a\,\rho\,\square^2\rho\Big)
\end{aligned}
\end{equation}
Adding the variation of a local term proportional to $(\square\rho)^2$, this becomes
\begin{equation}
\begin{aligned}
\delta_\s W=\int\sqrt{g}\,\s\,c\Big(R+{1\over16}\a\,R^2\,\rho^2
+{1\over4}\a\,R\,\rho\,\square\rho-{1\over2}\a\,R\,(\nabla\rho)^2+{1\over4}\a\,\square\rho\,\square\rho\Big)
\end{aligned}
\end{equation}
It does not contain $\tau$ at all. Nevertheless, the presence of $\tau$ was essential to have
a solution to the consistency condition for the Weyl-variation of the metric, which did contain $\tau$.
If, on the other hand, we had started with only $\tilde O$, which corresponds to setting $\a=0$, we would not have
have found any non-trivial consistent anomaly to ${\cal O}(\tau)$.

We now discuss the special role of $\tau$.
To begin with, a dimension four operator $\tilde O$ in $d=2$ with a linear coupling to energy-momentum tensors
is very special: obviously the coupling should start with two energy-momentum tensors.
Writing $\tilde O$ as the product of its left $ {\tilde O}_L$  and right  $ {\tilde O}_R$ components, we need a non-zero
correlator of  $ {\tilde O}_L$ with two left components of the energy-momentum tensor $T_L$.
In $d=2$, due to the infinite dimensional Virasoro symmetry,
the correlator is nonzero only if  $ {\tilde O}_L$ itself is proportional to $T_L$.
A similar argument holds for the right component and we conclude that $\tilde O$ has to
be identified with the famous $T_L T_R$ \cite{Zamo}.

We can see this happen quite generally. In $d=2$ the contribution to the ${\cal OO}$ OPE
from the vacuum representation, to which the energy-momentum tensor belongs, is fixed by the
Virasoro symmetry to
\begin{equation}
\begin{aligned}
{\cal O}(z,\bar z)\,{\cal O}(w,\bar w)&=\left({1\over(z-w)^3} +{3\over c}{1\over z-w}T_L(w)+\dots\right)
\left({1\over(\bar z-\bar w)^3} +{3\over c}{1\over \bar z-\bar w}T_R(\bar w)+\dots\right)\\
&=\dots+{9\over c^2} {1\over|z-w|^2}T_L(z)\bar T_R(\bar z)+\dots
\end{aligned}
\end{equation}
as one can explicitly verify for the free fermion model with $c={3\over2}$.
The operator $\tilde{\cal O} =T_LT_R$ indeed appears in the OPE of two ${\cal O}$ operators with a logarithmic coefficient.
The operator ${\cal O}$ has all its couplings determined by being a nonsingular product of energy-momentum tensors.
In our scheme, which preserves diffeoinvariance, it can be given an explicit diffeoinvariant definition as
\begin{equation}\label{cova}
\tilde{\cal O}= T_{\mu\nu}T^{\mu\nu}
\end{equation}
which differs from the original expressions by terms involving the trace $T_{\mu}^{\mu}$, which vanishes as an operator.
In correlators the two definitions differ therefore by local terms.
It is obvious that all the couplings of  $\tilde{\cal O}$ are determined by the couplings of the energy-momentum tensors
and therefore it cannot  generate  anomalies with independent universal coefficients. We can analyze this further
by studying the terms involving $\tilde{\cal O}$ to linear order in its coupling in the Osborn Equation.
To order $\rho^2$ there is a term in the equation which in leading order has the form:
\begin{equation}\label{danger}
\rho \log{{\Box \over \mu^2}} \rho\, \langle T_{\mu\nu}\rangle_1\langle T^{\mu\nu}\rangle_1
\end{equation}
originating in the product of the chirally factored correlators
$\langle {\cal O}_L  {\cal O}_L T_L \rangle \langle {\cal O}_R  {\cal O}_R T_R \rangle$. 
Here $\langle T_{\mu\nu}\rangle_1$  is the expansion of the nonlocal energy-momentum tensor defined from the 
non-local Polyakov action, to first order around the flat metric, i.e. 
\begin{equation}\label{rrTT}
\langle T_{\mu\nu}\rangle_1\propto \left(\eta_{\mu\nu}-{\p_\mu\p_\nu\over\square}\right)
\left(\p^\a\p^\b-\eta^{\a\b}\square\right)h_{\a\b}
\end{equation}
The term in \eqref{danger} has an explicit logarithmic singularity and when in higher order more energy-momentum tensors
are correlated, i.e. \eqref{rrTT}  is evaluated in the background of a generic metric $g_{\mu\nu}$. 
When acted upon by the standard  metric variation the logarithmic singularity is eliminated
but the powerlike singularities remain. In order to cancel them an explicit (non-local) term linear in $\tau$ should be introduced
\begin{equation}\label{saviour}
\int d^2 x\sqrt{g}\,\tau(x)\langle T_{\mu\nu}T^{\mu\nu} \rangle_g
\end{equation}
where the matrix element of the energy-momentum tensors has to be evaluated in a generic background $g_{\mu\nu}$.
This shows that without the explicit coupling linear in $\tau$, the Osborn Equation is not consistent to order $\rho^2$ and
therefore we should expect that this is reflected on the metric beta-function. Indeed we found that consistency  
requires that it contains both $\rho$ and $\tau$. 
Now we discuss the second puzzle raised before, i.e. why the Weyl anomaly does not contain $\tau$-dependent terms.
After the  addition of \eqref{saviour} we should consider it together with the pure $g_{\mu\nu}$
dependent (Polyakov) term, i.e.
\begin{equation}\label{partial}
\tilde W = -{c\over4} \int d^2 x \sqrt{g}\, R {1 \over \Box}R
+\int d^2 x\sqrt{g}\,\tau(x)\langle T_{\mu\nu}\rangle_g \langle T^{\mu\nu}\rangle_g
\end{equation}
For the evaluation of the energy-momentum matrix element we used the fact that in a chiral scheme the  correlator
is factored into a ``left" and ``right" part, and therefore in a general background in the diff-invariant scheme it can be also
evaluated in a factorized form.  The result is valid modulo local terms. It is clear that the second term does not contain
new universal information. More explicitly we can show to order  $\tau$  that \eqref{partial} obeys a homogenous
Osborn Equation. The Weyl variation of $\tilde W$ at order $\tau$ is
\begin{equation}\label{var}
\delta_{\s}\tilde W=\int d^2 x \sqrt{g}\,T^{\mu\nu}\big(\b ^g _{\mu\nu} +2\, \tau\, \delta_{\s} T_{\mu\nu}\big)
\end{equation}
where $\b^g$ is to be determined and the energy-momentum tensor are determined from the Polyakov action.
Using
\begin{equation}
\d_\s T_{\mu\nu}={c\over4}\big(\nabla_\mu\nabla_\nu\s-g_{\mu\nu}\,\square\s\big)
\end{equation}
the second term leads in \eqref{var} to the term $\int\sqrt{g}\,\tau\,R\,\square\s$, which is cohomologically
trivial, being the Weyl variation of $\int\sqrt{g}\,\tau\,R^2$. The first term is cohomologically non-trivial and would
lead to a non-local expression for $\d_\s W$. Requiring this to vanish determines $\b^g$
\begin{equation}\label{fake}
\b ^g _{\mu\nu} = -{c\over2} \tau\,\nabla_\mu\nabla_\nu\s
\end{equation}
in agreement with \eqref{dg2}, after rescaling $\tau$ appropriately.

We see that the $\tau$-dependent part of the metric beta-function is entirely determined by the Polyakov action
and does not reflect any new universal coefficient of a logarithm.
As a consequence there is no contribution to the anomaly.
This is a very special feature of $d=2$ related
to the fact that the $\Delta=4$ operator is the ubiquitous $T\bar T$ \cite{Zamo}. 
There is of course an ambiguity of local terms which could be added to the generating functional.

\section{Conclusions}

The main result of this paper is that integer dimension 
irrelevant operators change in a significant fashion the structure of the Osborn Equation.
The presence of their sources in the generating functional requires 
a non-trivial beta-function for the metric, i.e. an addition to its classical transformation rule under Weyl rescaling 
which depends on the irrelevant sources. 
To any given finite order in the number of irrelevant sources the equation stays consistent and a local Weyl anomaly can be
defined. It contains the information about  the normalization of universal single logarithms in various correlators.

We now summarize the pattern for beta-functions and Weyl anomalies which we find in any even dimension
$d$ ($d=2$ having the special features discussed in Section 4):

\begin{itemize}

\item[a.] Starting with an integer dimension $\Delta>d$ operator  ${\cal O}$ with source $\rho$, there is a logarithmic
singularity  in the two-point funtion $\langle{\cal O\,O}\rangle$ and a double logarithm in
the correlator $\langle T {\cal O}{\cal O}\rangle$.

\item[b.] As  a consequence there is a metric beta function to order $\rho^2$.

\item[c.] The metric beta-function implies a deformation of the Weyl anomaly generated
by the correlator of two energy-momentum tensors. The same anomaly contains the information about the
two point function of ${\cal O}$, thus  generalizing the  type B  anomalies of marginal and relevant operators.

\item[d.] The other Weyl anomalies involving energy-momentum tensors are deformed to
order $\rho^2$ by the presence of the metric beta-function. They are related to decompositions of the two
$\langle T {\cal O O\rangle}$ energy-momentum correlators, where the two ${\cal O}$'s couple
through the energy-momentum tensor conformal block.

\item[e.] To order $\rho^2$ there are beta-functions generated for any integer dimensional $\tilde{\cal O}$
operator with dimension $\tilde\Delta \leq (2 \Delta -d)$ and coupling $\tau$. These beta-functions deform
the type B anomalies which are linear in $\tau$.
To ${\cal O}(\rho^2)$  the deformed Weyl anomalies are related to
correlators of  two ${\cal O}$'s and energy-momentum tensors, where the two  ${\cal O}$'s
couple to the energy-momentum tensors through the $\tilde{\cal O}$ conformal block.

\end{itemize}

An application of the formalism discussed in this paper is the study of correlators of elements of the chiral ring in
${\cal N}=2$ superconformal theories in four dimensions \cite{ChiralRing1,ChiralRing2}.
The chiral ring contains  irrelevant operators and their correlators can be studied using the methods of this
paper and the anomaly equations generalized to a supersymmetric set up.
The powers of curvatures present in the anomaly
formulae define the operator mixings involved.
Moreover the anomaly coefficients can depend on constant moduli.
We are currently studying this generalization in more detail.

A natural question is the physical meaning of the metric beta-function.
In a CFT the metric acts  as a source for the energy-momentum tensor and the beta-function is just a
convenient way to organize the recursion relations connecting the various powers of the logarithms.
When however the irrelevant couplings are turned on as true massive deformations of the CFT, the metric
beta-function indicates that the metric considered as a coupling also starts running.
A natural interpretation of the metric as a coupling is when the theory is formulated on a curved manifold:
the parameters characterizing the geometry of the manifold can be considered to be coupling constants which
multiply the terms in the Lagrangian involving energy-momentum tensors.
In particular when the theory is formulated on $S_d$ the radius becomes a coupling constant in the above sense.
We reach the strange conclusion that the radius of the sphere on which the theory is formulated
has a nontrivial renormalization group transformation. The scale dependent contribution
to the partition function following from the transformation of the radius enters through the anomaly:
for the $\Delta=5$ theory in $d=4$ the $``c"$ anomaly for constant $\rho$ vanishes on
the sphere but not the $``a"$ anomaly,  such that such a new type of dependence is possible in principle.
The strange feature mentioned above depends of course on turning on the irrelevant perturbation which
is not consistent outside a finite order of perturbation theory.

The structure of the anomalies corresponding to terms with multiple logarithms in the generating functional can be
understood by choosing a decomposition of the correlators in conformal blocks
where the two irrelevant operators couple first to an integer dimensional block.
It is an interesting question how the logarithmic structures are reproduced by other
decompositions related by crossing.

\noindent
{\bf Acknowledgements:}
We thank E. Brehm, H. Godazgar, R. Manvelyan and E. Skvortsov for discussions.
Some of the calculations were performed
with the help of the xAct collection of Mathematica packages.
This work was supported in part by I-CORE program of the Planning and Budgeting Committee and the
Israel Science Foundation (grant number 1937/12) and by
GIF -- the German-Israeli Foundation for Scientific Research and Development (grant number 1265).

\end{document}